\documentclass[review]{elsarticle}

\usepackage{hyperref}

\usepackage{tikz-feynman}
\usepackage{amsmath,amssymb}
\usepackage{siunitx}
\usepackage[T1]{fontenc}
\usepackage[utf8]{inputenc}

\journal{Physics Letters B}

\newcommand{\diff}[2][]{%
\,\mathit{d}^{#1}#2\,%
}
\newcommand{\Int}{\int}
\newcommand{\vect}[1]{\mathbf{#1}}
\newcommand{\figref}[1]{Fig.~\ref{#1}}
\begin{document}
\begin{frontmatter}

\title{Coulomb corrections to the bremsstrahlung and electron pair production
  cross section of high-energy muons on extended nuclei}
\author[TUDo]{A. Sandrock}
\author[TUDo]{W. Rhode}
\address[TUDo]{Department of Physics, TU Dortmund University, D-44221 Dortmund,
 Germany}

\begin{abstract}
The energy reconstruction of high-energy muons depends on the energy loss
characteristics. Accurate knowledge of the cross sections of the energy loss
processes is necessary for precise measurements of the energy spectrum of muons
and muon-induced neutrinos.

The cross sections of the two most dominant processes, electron pair production
and bremsstrahlung are calculated exactly in the coupling parameter $Z\alpha$ to
the electromagnetic field of a nucleus for realistic extended screened nuclei.
An analytical parametrization of the mass and nuclear-charge dependence of the
cross section is given.
\end{abstract}


\end{frontmatter}

\section{Introduction}
The energy reconstruction of high-energy muons is a central task in cosmic-ray
and neutrino astronomy experiments. The energy is reconstructed based on the
energy loss. The dominant energy loss processes of high-energy muons are
electron pair production \cite{KokoulinPetrukhin1,KokoulinPetrukhin2,%
Kelner-atomic}, bremsstrahlung \cite{KKP,KKP-atomic}, and
inelastic nuclear interaction \cite{ALLM91,ALLM97,BezrukovBugaev}. The
energy is lost stochastically, however the average energy lost per distance
is well described by a quasi-linear function of the energy
\begin{equation}
-\frac{dE}{dx} = a(E) + b(E) E,
\end{equation}
where $a, b$ only weakly depend on the energy. The uncertainties of the
cross sections influence the systematic uncertainties of the energy
reconstruction \cite{KokoulinUncertainties}.

Currently used parametrizations of the pair production and bremsstrahlung
cross section are calculations in the Born approximation which take into
account the screening of the nucleus by atomic electrons
\cite{KokoulinPetrukhin1,PetrukhinShestakov}, the effect of the extended
nucleus \cite{KokoulinPetrukhin2,PetrukhinShestakov,KKP} and the contribution
of atomic electrons as target particles \cite{Kelner-atomic,KKP-atomic}.

The effect of higher-order corrections in the nuclear coupling constant $Z
\alpha$, where $Z$ is the nuclear charge and $\alpha$ the fine structure
constant, so-called Coulomb corrections, has been considered for pair
production in \cite{Ivanov,Melnikov} for a point-like nucleus. However, in
\cite{Ivanov}, it was pointed out that the effect of a form factor can be
sizeable. In this article, the corrections are calculated for a realistic 
nuclear charge distribution. This process has been considered recently in
\cite{KrachkovPair} in the quasiclassical approximation, neglecting the nuclear
form factor, but taking into account higher-order corrections to the interaction
of the initial charged particle with the nucleus, which will be neglected in
the following, because only muons are considered as initial charged particle,
while in \cite{KrachkovPair} emphasis was put also on ions as initial particles.
In this article, the correction to the spectrum of secondary particles and also
the average energy loss of the muon is calculated.

The effect of Coulomb corrections on the bremsstrahlung cross section has been
calculated in \cite{DaviesBetheMaximon} for electrons and was shown to be
large (from $\sim 1\%$ for medium nuclei such as iron to $\sim 10\%$ for heavy
nuclei such as uranium). In \cite{ABB2} Coulomb corrections for muon
bremsstrahlung on extended nuclei were calculated in the approximation of a
homogeneously charged sphere for the nuclear charge density and were found to
be small ($\lesssim 0.5\%$). This process was also considered more recently in
\cite{KrachkovBrems}, in whose approximation the correction for muons vanishes
identically, independently of the form of the nuclear potential.

\section[Corrections for point-like nucleus]{Higher order corrections in
  $Z\alpha$ for a point-like nucleus}
The coupling to the field of the nucleus is governed by the coupling constant
$\nu = Z \alpha$, which for high $Z$ can achieve values which are not very small
compared to 1, and therefore should be treated non-perturbatively. The first
work on this subject was \cite{DaviesBetheMaximon}, where the corrections to
the cross sections of bremsstrahlung and pair production by real photons were
calculated using wave functions which are approximate solutions of the Dirac
equation for a Coulomb field. The final result is in the case of pair
production by a real photon the expression
\begin{equation}
\begin{split}
\frac{\diff\sigma}{\diff x} &= \frac{4}{3} Z^2 \alpha r_e^2 \{1 + 2 [x^2 +
  (1 - x)^2]\} \left(\ln \frac{2 x (1 - x) \omega}{m} - \frac{1}{2}
  - f(Z \alpha)\right),\\
f(\nu) &= \nu^2 \sum_{n = 1}^\infty \frac{1}{n (n^2 + \nu^2)},
\end{split}
\end{equation}
where $x = \epsilon_+/\omega$ is the ratio of the initial photon energy $\omega$
and the positron energy $\epsilon_+$. It is very difficult to extend this
treatment to a more realistic description of the nucleus as the wave functions
would have to be determined for the given potential. The term $\ln [2 x (1 - x)
\omega/m] - \frac{1}{2}$ arising from the leading order calculation is called
the main logarithm in the following\footnote{For an atomic field different from
the Coulomb case, the main logarithm changes also.}, the term
$f(Z \alpha)$ arising from higher-order corrections in the coupling to the
nuclear field is called Coulomb correction.

In \cite{Ivanov}, this result was obtained again in a much simpler way by
resummation of the perturbation series. Moreover, the approach in \cite{Ivanov}
allows for the inclusion of realistic atomic and nuclear form factors. In
\cite{Bakmaev} this approach was applied to the problem of electron-positron
photoproduction in the field of a screened nucleus. In \cite{Melnikov}, the
results of \cite{Ivanov} were used to calculate the Coulomb correction to the
pair production cross section by high-energy muons on a pointlike nucleus.
First, the calculations of \cite{Ivanov} are briefly reviewed and then this
treatment is extended to calculate the corrections for a screened extended
nucleus which allows also to determine the Coulomb correction to muon bremsstrahlung.

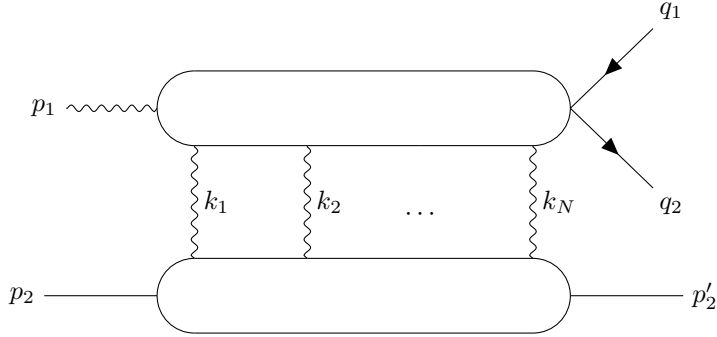
\begin{figure}
\begin{tikzpicture}
  \begin{feynman}
    \vertex (p1) {\(p_1\)};
    \vertex [right=of p1] (qq);
    \vertex [below right=2em of qq] (k1);
    \vertex [right=of k1] (k2);
    \vertex [right=of k2] (k);
    \vertex [below=2em of k] (dots) {\(\cdots\)};
    \vertex [right=of k] (kN);
    \vertex [above right=2em of kN] (QQ);
    \vertex [above right=of QQ] (q1) {\(q_1\)};
    \vertex [below right=of QQ] (q2) {\(q_2\)};
    \vertex [above right=2em of qq] (l1);
    \vertex [above left=2em of QQ] (l2);
    \vertex [below=of k1] (K1);
    \vertex [below=of k2] (K2);
    \vertex [below=of kN] (KN);
    \vertex [below left=2em of K1] (pp);
    \vertex [below right=2em of pp] (L1);
    \vertex [below right=2em of KN] (PP);
    \vertex [below left=2em of PP] (L2);
    \vertex [left=of pp] (p2) {\(p_2\)};
    \vertex [right=of PP] (p2p) {\(p_2'\)};

    \diagram*{
      (p1) -- [boson] (qq),
      (qq) -- [quarter left] (l1) -- (l2) -- [quarter left] (QQ) --
        [quarter left] (kN) -- (k) -- (k2) -- (k1) -- [quarter left] (qq),
      (k1) -- [boson, edge label=\(k_1\)] (K1),
      (k2) -- [boson, edge label=\(k_2\)] (K2),
      (kN) -- [boson, edge label=\(k_N\)] (KN),
      (K1) -- (K2) -- (KN) -- [quarter left] (PP) -- [quarter left] (L2) --
        (L1) -- [quarter left] (pp) -- [quarter left] (K1),
      (pp) -- (p2),
      (PP) -- (p2p),
      (q1) -- [fermion] (QQ) -- [fermion] (q2),
    };
  \end{feynman}
\end{tikzpicture}
\caption{Diagram of the pair production process with $N$ photons exchanged with
  the nucleus.}
\label{Coulomb-diag}
\end{figure}
The main contribution to the cross section arises from small scattering angles,
therefore the momenta (cf. \figref{Coulomb-diag}) are expressed in Sudakov
variables \cite{Sudakov}
\begin{equation}
\begin{split}
k_i &= \alpha_i \tilde p_1 + \beta_i \tilde p_2 +  k_{i\perp},\\
q_i &= x_i \tilde p_1 + y_i \tilde p_2 + q_{i\perp},\\
\end{split}
\end{equation}
where $\tilde p_1 = p_1 + (Q^2/s) p_2$, $\tilde p_2 = p_2 - (m^2/s) p_1$ are
almost light-like vectors, $Q^2 = -p_2^2$ is the virtuality of the photon and
$s = 2 p_1 p_2 \gg Q^2, m^2$ is the center of momentum energy. As a
simplification, the mass of the nucleus and of the produced lepton are set
equal $p_2^2 = q_1^2 = q_2^2 = m^2$. The mass of the nucleus does not enter the
final result, where its mass is considered infinite. Denoting again by $x$ the
fraction of the energy of the initial photon $\omega$ which is
transferred to the antilepton, 
\begin{align*}
x_1 &= x, & x_2 &= 1 - x,\\
y_1 &= \frac{m^2 + \vect q_1^2}{x s}, & y_2 &= \frac{m^2 + \vect q_2^2}
  {(1 - x) s}
\end{align*}
with $\vect q_i^2 = -q_{i\perp}^2$. The amplitude for the diagram with $N$
exchanged photons is given in the impact representation \cite{Lipatov} by
\begin{equation}
\mathcal{M}_N = \frac{8 \pi^2 s (-i)^{N-1}}{N!} \Int \prod_{i=1}^N
  \left(\frac{\diff[2]{\vect k_i}}{(2 \pi)^2 \vect k_i^2}\right)
  \delta \left(\sum_{j = 1}^N \vect k_j - \vect q_1 - \vect q_2\right)
  J_{\gamma \rightarrow \ell \bar\ell}^N J_A^N.
\label{eq-5.5}
\end{equation}
The impact factors are given by
\begin{align}
J_{\gamma \rightarrow \ell \bar \ell}^N &= \Int \prod_{i = 1}^{N - 1}
  \left(\frac{\diff{(\beta_i s)}}{2 \pi i}\right) (iA)_{\mu_1 \dots \mu_N}
  \frac{\tilde p_2^{\mu_1} \cdots \tilde p_2^{\mu_N}}{s^N},\\
J_A^N &= \Int \prod_{i = 1}^{N - 1} \left(\frac{\diff{(\alpha_i s)}}{2 \pi i}
  \right) (iB)_{\mu_1 \dots \mu_N} \frac{\tilde p_1^{\mu_1} \cdots
  \tilde p_1^{\mu_N}}{s^N}.
\end{align}
where $(iA)_{\mu_1 \dots \mu_N}$ is the amplitude corresponding to the upper
part of the diagram in \figref{Coulomb-diag} and $(iB)_{\mu_1 \dots \mu_N}$ to
the lower part.

For an infinitely heavy point nucleus, the impact factor is
given by
\begin{equation}
J_A^N = i (-1)^N (e Z)^N.
\end{equation}
Accounting for an extended nucleus can be either carried out by modifying the
impact factor or equivalently by modifying the Coulomb propagator
$1/\vect k_i^2$ in \eqref{eq-5.5}.

The impact factors for the lepton part of the diagram can be determined by a
recurrence relation. The impact factor for one exchanged photon is given by
\begin{equation}
J_{\gamma \rightarrow \ell \bar\ell}^1(\vect q_1, \vect q_2) = i e^2 \bar u_1
  [m \hat{\vect e} S^1 - 2 x (\vect T^1 \vect e) - \hat{\vect T}^1
  {\hat{\vect e}}] \frac{\hat{\tilde {p}}_2}{s} u_2
\end{equation}
for a transversely polarized photon with polarization vector $\vect e$ and
for a longitudinally polarized photon by
\begin{equation}
J_{\gamma \rightarrow \ell \bar\ell}^1 (\vect q_1, \vect q_2) = -i e^2
  \sqrt{Q^2} x (1 - x) S^1 (\vect q_1, \vect q_2) \bar u_1
  \frac{\hat{\tilde {p}}_2}{s} u_2,
\end{equation}
where
\begin{align}
S^1 \equiv S^1 (\vect q_1, \vect q_2) &= \frac{1}{\mu^2 + \vect q_1^2}
  - \frac{1}{\mu^2 + \vect q_2^2},\\
\vect T^1 \equiv \vect T^1 (\vect q_1, \vect q_2) &= \frac{\vect q_1}
  {\mu^2 + \vect q_1^2} + \frac{\vect q_2}{\mu^2 + \vect q_2^2},\\
\mu^2 &= m^2 + Q^2 x (1 - x).
\end{align}

The scalar $S^N$ and vector $\vect T^N$ structures are related by the recurrence
relations
\begin{align}
S^N (\vect q_1, \vect q_2, \vect k_N) &= S^{N - 1}(\vect q_1, \vect q_2
  - \vect k_N) - S^{N - 1} (\vect q_1 - \vect k_N, \vect q_2),\\
\vect T^N (\vect q_1, \vect q_2, \vect k_N) &= \vect T^{N - 1} (\vect q_1,
  \vect q_2 - \vect k_N) - \vect T^{N - 1} (\vect q_1 - \vect k_N, \vect q_2),
\end{align}
because due to Bose symmetry the $N$-th $t$-channel photon can be considered as
the last one attached to the lepton line, from which the relations follow 
immediately \cite{Ivanov-mail}. The dependence on the other $t$-channel photon
momenta $\vect k_1, \dots \vect k_{N - 1}$ is omitted for clarity. The integral
over the $t$-channel momenta
\begin{equation}
J_e^N (\vect q_1, \vect q_2) = \Int \prod_{i = 1}^N \frac{\diff[2]{\vect k_i}}
  {\vect k_i^2} F(\vect k_i) \delta \left( \sum_{j = 1}^N \vect k_j
  - \vect q\right) S^N
\end{equation}
with the formfactor $F(\vect k)$ can be recast using the recurrence relations as
\begin{equation}
J_S^N (\vect q_1, \vect q_2) = \Int \frac{\diff[2]{\vect k}}{\vect k^2}
  F(\vect k) [J_S^{N - 1} (\vect q_1, \vect q_2 - \vect k) - J_S^{N - 1}
  (\vect q_1 - \vect k, \vect q_2)]
\end{equation}
such that for the Fourier transform of $J_S^N(\vect q_1, \vect q_2)$
\begin{equation}
j_S^N(\vect r_1, \vect r_2) = \frac{1}{(2 \pi)^2} \Int e^{i \vect q_1 \vect r_1
  + \vect q_2 \vect r_2} J_S^N (\vect q_1, \vect q_2) \diff[2]{\vect q_1}
  \diff[2]{\vect q_2}
\end{equation}
the recurrence relation assumes the form
\begin{equation}
\begin{split}
j_S^N(\vect r_1, \vect r_2) &= j_S^{N - 1} (\vect r_1, \vect r_2) \pi
  \phi(\vect r_1, \vect r_2)\\
\phi (\vect r_1, \vect r_2) &= \frac{1}{\pi} \Int \frac{\diff[2]{\vect k}}
  {\vect k^2} (e^{i \vect k \vect r_2} - e^{i \vect k \vect r_1}).
\end{split}
\end{equation}
Using the Fourier transform of $J_S^1$
\begin{equation}
j_S^1 (\vect r_1, \vect r_2) = \frac{1}{2} K_0 (\mu |\vect r_1 - \vect r_2|)
  \phi(\vect r_1, \vect r_2),
\end{equation}
 the total impact factor to all orders, inverting the Fourier
transform, is given by
\begin{equation}
\begin{split}
J_S (\vect q_1, \vect q_2) &= \frac{i}{(2 \pi)^2 2 \nu} \Int \diff[2]{\vect r_1}
  \diff[2]{\vect r_2} e^{-i \vect q_1 \vect r_1 - i \vect q_2 \vect r_2}\\
&\times K_0 (\mu |\vect r_1 - \vect r_2|)
   [e^{-i \nu \phi(\vect r_1, \vect r_2)} - 1],
\end{split}
\end{equation}
and analogously for the vector structure by
\begin{equation}
\begin{split}
\vect J_T (\vect q_1, \vect q_2) &= \frac{-1}{(2 \pi)^2 2 \nu}
  \Int \diff[2]{\vect r_1} \diff[2]{\vect r_2}
  e^{-i \vect q_1 \vect r_1 - i \vect q_2 \vect r_2}\\
&\times \frac{\mu (\vect r_1 - \vect r_2)}{|\vect r_1 - \vect r_2|}
  K_1 (\mu |\vect r_1 - \vect r_2|)
  [e^{-i \nu \phi(\vect r_1, \vect r_2)} - 1].
\end{split}
\end{equation}
Here $K_\nu(z)$ is the modified Bessel function. To obtain the amplitude out of
the impact factor, according to \cite{Suura} the impact factor is multiplied
by a universal phase factor and the amplitude is given by
\begin{equation}
\begin{split}
\mathcal{M} &= 8 \pi e \nu s \left(\frac{x}{1 - x}\right)^{-i\nu}
  \bar u_1 \left\{m \hat{\vect e} J_S (\vect q_1, \vect q_2) \right.\\
&- \left. 2 x [\vect J_T (\vect q_1, \vect q_2) \vect e] - \hat{\vect J}_T
  (\vect q_1, \vect q_2) \hat{\vect e}\right\} \frac{\hat{\tilde {p}}_2}{s} u_2
\end{split}
\end{equation}
for a transversely polarized incident photon and by
\begin{equation}
\mathcal{M} = -16 \pi e \nu s \left(\frac{x}{1 - x}\right)^{-i \nu} \sqrt{Q^2}
  x (1 - x) \bar u_1 J_S (\vect q_1, \vect q_2) \frac{\hat{\tilde {p}}_2}{s} u_2
\end{equation}
for a longitudinally polarized incident photon.

The total cross section is obtained by integration over the transversal momenta
$\vect q_1, \vect q_2$ and the energy fraction $x$ as
\begin{equation}
\diff\sigma = \frac{2 \nu^2 \alpha}{\pi^2} \{m^2 |J_S|^2 + |\vect J_T|^2 [x^2 +
  (1 - x)^2] \} \diff x \diff[2]{\vect q_1} \diff[2]{\vect q_2}
\end{equation}
for the transversely polarized photon, summed over all polarization states. To
obtain the Coulomb correction to the Born cross section, the Born approximation
cross section has to be subtracted. Therefore the correction $\diff{\sigma_2}$,
for $\diff\sigma = \diff{\sigma_1} + \diff{\sigma_2}$ with $\diff{\sigma_1}$ the
Born approximation cross section, is given for the transversely and
longitudinally polarized photon by
\begin{align}
\frac{\diff{\sigma_2^T}}{\diff x} &= \frac{2 \alpha \nu^2}{\pi^2} \{m^2 A_1
  + [x^2 + (1 - x)^2] A_2\},\\
\frac{\diff{\sigma_2^S}}{\diff x} &= \frac{2 \alpha \nu^2}{\pi^2} 4 Q^2 x^2
  (1 - x)^2 A_1,
\end{align}
respectively, where
\begin{align}
A_1 &= \Int \diff[2]{\vect q_1} \diff[2]{\vect q_2} (|J_S|^2 - |J_S^1|^2),\\
A_2 &= \Int \diff[2]{\vect q_1} \diff[2]{\vect q_2} (|\vect J_T|^2
  - |\vect J_T^1|^2).
\end{align}

\section{Higher-order corrections for an extended screened nucleus}
In the calculation of corrections in a Coulomb field, the expressions for
$A_1, A_2$ contain terms which diverge and have to be regularized, which leads
to not well-defined expressions when attempting a numerical integration. As
pointed out by \cite{Bakmaev}, the divergences are removed when screening is
taken into account.

Using the form factor \cite{Tsai}
\begin{align*}
F(k) &= F_n(k) - F_e(k),\\
F_n(k) &= \left(1 + \frac{a^2 k^2}{12}\right)^{-2}, &
  a &= (\num{0.58} + \num{0.82} A^{1/3}) \SI{5.07}{GeV^{-1}}\\
F_e(k) &= \frac{1}{1 + b^2 k^2}, & b &= \frac{\num{184.15} Z^{-1/3}}
  {m_e \sqrt{e}},
\end{align*}
$\phi$ is given by
\begin{equation}
\begin{split}
\phi(\vect r_1, \vect r_2) &= \frac{1}{\pi} \int \frac{\diff[2]{\vect k}}
  {\vect k^2} F(|\vect k|) (e^{i \vect k \vect r_2} - e^{i \vect k \vect r_1})
  \\
&= 2 [K_0 (\Lambda_e r_2) - K_0 (\Lambda_e r_1)]
  + 2 [K_0 (\Lambda_n r_1) - K_0 (\Lambda_n r_2)]\\
&+ \Lambda_n r_1 K_1 (\Lambda_n r_1) - \Lambda_n r_2 K_1 (\Lambda_n r_2),\\
\Lambda_e &= \frac{1}{b}, \quad \Lambda_n = \frac{\sqrt{12}}{a}.
\end{split}
\end{equation}
The quantities $A_1, A_2$ are given by the expressions
\begin{align}
A_1 &= \frac{\pi}{2 \nu^2 \mu^4}
  \Int_0^\infty dx \Int_0^\infty dR \Int_0^{2 \pi} \diff\theta
  x^3 {K_0}^2(x) \{2 - 2 \cos (\nu \phi_{12})- \nu^2 \phi_{12}^2 \},\\
A_2 &= \frac{\pi}{2 \nu^2 \mu^2}
  \Int_0^\infty dx \Int_0^\infty dR \Int_0^{2 \pi} \diff\theta
  x^3 {K_1}^2(x) \{2 - 2 \cos (\nu \phi_{12})- \nu^2 \phi_{12}^2 \},\\
\phi_{12} &= \phi\left(\frac{x R}{\mu},
  \frac{x \sqrt{R^2 + 1 - 2 R \cos\theta}}{\mu} \right).
\end{align}

In the case of a Coulomb field, $\phi = \ln (\vect r_1^2/ \vect r_2^2)$ and
$A_1, A_2$ assume the values
\begin{align*}
A_1^\text{C} &= -\frac{2 \pi^2}{3 \mu^4} f(\nu),\\
A_2^\text{C} &= -\frac{4 \pi^2}{3 \mu^2} f(\nu),\\
f(\nu) &= \frac{1}{2} \{ \Psi (1 - i \nu) + \Psi (1 + i \nu) - 2 \Psi(1)\}\\
  &= \nu^2 \sum_{n = 1}^\infty \frac{1}{n (n^2 + \nu^2)}.
\end{align*}
When realistic form factors are employed, it is no longer possible to evaluate
the Coulomb corrections in closed form. The numerical results can be
approximated by
\begin{align}
A_1 &= A_1^\text{C} g_1 (\mu/\si{MeV}, \nu), \quad
  A_2 = A_2^\text{C} g_2 (\mu/\si{MeV}, \nu),\\
g_i (x, \nu) &= \frac{a_i(\nu) + b_i(\nu) x}{1 + c_i(\nu) x + d_i(\nu) x^2},
\end{align}
where $a_i, b_i, c_i, d_i$ are approximately cubic polynomials for $Z > 5$
\begin{align*}
a_1(\nu) &= \num{1.0026} - \num{2.2789e-2} \nu + \num{2.9437e-2} \nu^2
  - \num{4.1536e-2} \nu^3,\\
b_1(\nu) &= \num{1.9465e-2} - \num{7.7063e-2} \nu + \num{1.9979e-1} \nu^2
  - \num{1.4107e-1} \nu^3,\\
c_1(\nu) &= \num{3.6785e-2} + \num{5.4466e-2} \nu - \num{9.2971e-2} \nu^2
  + \num{2.7357e-2} \nu^3,\\
d_1(\nu) &= \num{9.9382e-4} + \num{2.4601e-3} \nu + \num{2.6733e-3} \nu^2
  - \num{2.8198e-3} \nu^3;\\
a_2(\nu) &= \num{1.0046} - \num{1.9267e-2} \nu + \num{4.5255e-2} \nu^2
  - \num{5.1603e-2} \nu^3,\\
b_2(\nu) &= \num{8.8223e-3} - \num{5.2931e-2} \nu + \num{1.4854e-1} \nu^2
  - \num{1.0764e-1} \nu^3,\\
c_2(\nu) &= \num{3.7141e-2} + \num{1.2897e-1} \nu - \num{2.2677e-1} \nu^2
  + \num{1.0776e-1} \nu^3,\\
d_2(\nu) &= \num{7.1144e-4} + \num{1.7710e-3} \nu + \num{5.0240e-3} \nu^2
  - \num{4.5527e-3} \nu^3.
\end{align*}
Since the correction is small for low $Z$, it is possible to use this
parametrization for all $Z$.

\begin{figure}
\includegraphics[width=\textwidth]{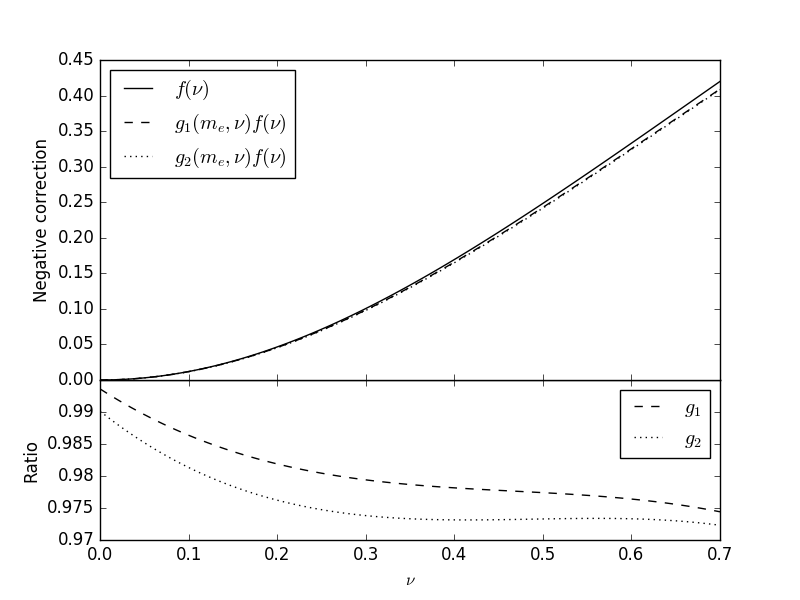}
\caption{Correction to the main logarithm of bremsstrahlung and photoproduction
for electrons due to Coulomb corrections for a Coulomb field and a screened
nucleus. Shown are the correction $f(\nu)$ for a Coulomb field (solid line) and
the corrections which account for the screened nucleus $g_1(m_e, \nu) f(\nu)$
(dashed line), $g_2(m_e, \nu) f(\nu)$ (dotted line).}
\label{coulomb}
\end{figure}

\begin{figure}
\includegraphics[width=\textwidth]{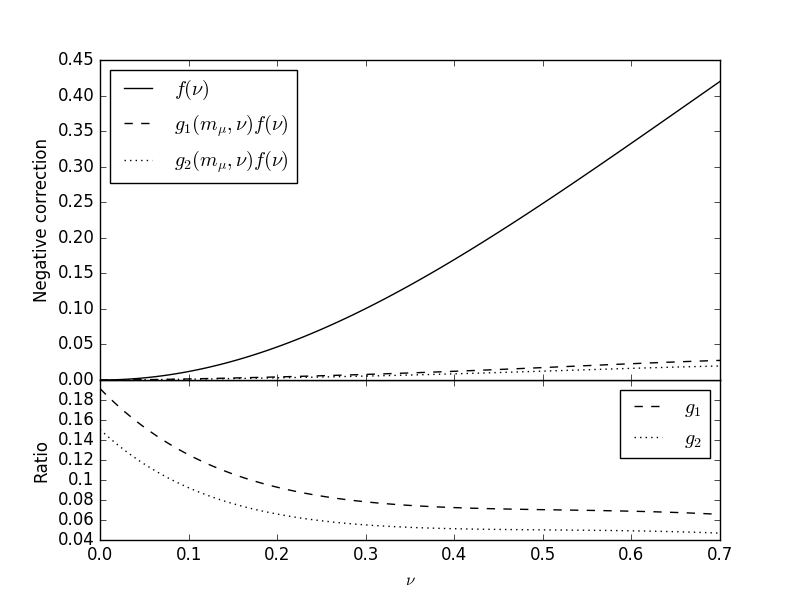}
\caption{Correction to the main logarithm of bremsstrahlung and photoproduction
for muons due to Coulomb corrections for a Coulomb field and a screened
nucleus. Shown are the correction $f(\nu)$ for a Coulomb field (solid line) and
the corrections which account for the screened nucleus $g_1(m_\mu, \nu) f(\nu)$
(dashed line), $g_2(m_\mu, \nu) f(\nu)$ (dotted line).}
\label{coulomb-muon}
\end{figure}

This Coulomb correction for the virtual photon pair production can be used to
calculate several cross sections. Setting $Q^2 = 0, \mu^2 = m^2$, one obtains
the corrections for real photoproduction of particles with mass $m$ on a
screened extended nucleus. The numerical examples show that for electrons, the
result of \cite{DaviesBetheMaximon} is reproduced with a small correction
for heavy nuclei (see \figref{coulomb}), while for muons the correction due to
multiphoton exchange is very small (see \figref{coulomb-muon}). Since the main
logarithm assumes the value $\ln [B Z^{-1/3}(m_\mu/m_e)] - \ln (1.54 A^{0.27})$
with $B \simeq 183$ \cite{KKP} in the full-screening limit, the correction to
the energy loss spectrum due to Coulomb corrections is negligible with very
high accuracy
\begin{equation*}
\max_{Z} \frac{f(\nu) g_{1,2}(m_\mu, \nu)}{\ln \left(B\frac{m_\mu}{m_e}
  Z^{-1/3}\right) - \ln (1.54 A^{0.27})} < 0.004.
\end{equation*}
The small influence of the nuclear form factor on electrons and
the smallness of the corrections for heavy particles was already observed by
\cite{Ivanov} in the limiting cases $\Lambda \gg m$ for electrons and
$\Lambda \ll m$ for muons, using a nuclear form factor
\begin{equation}
F(k) = \frac{\Lambda^2}{\Lambda^2 + k^2}.
\end{equation}
From the corrections to the real photoproduction cross section,
the corrections to the bremsstrahlung cross section are obtained via the
substitution rules
$\epsilon_+ \rightarrow -\epsilon_1, \epsilon_- \rightarrow \epsilon_2,
\omega \rightarrow -\omega, \diff\sigma \rightarrow (\omega^2 \diff\omega/
\epsilon_+^2 \diff{\epsilon_+}) \diff\sigma$, where $x = \epsilon_+/
\omega$ (e.\,g., \cite{Olsen55}).
Again, the classical result for electrons is obtained, that the
function $f(\nu)$ is subtracted from the main logarithm, and it is observed
that the correction for muon bremsstrahlung is small, as was found in
\cite{ABB2} for a simplified nuclear form factor.

\begin{figure}
\includegraphics[width=\textwidth]{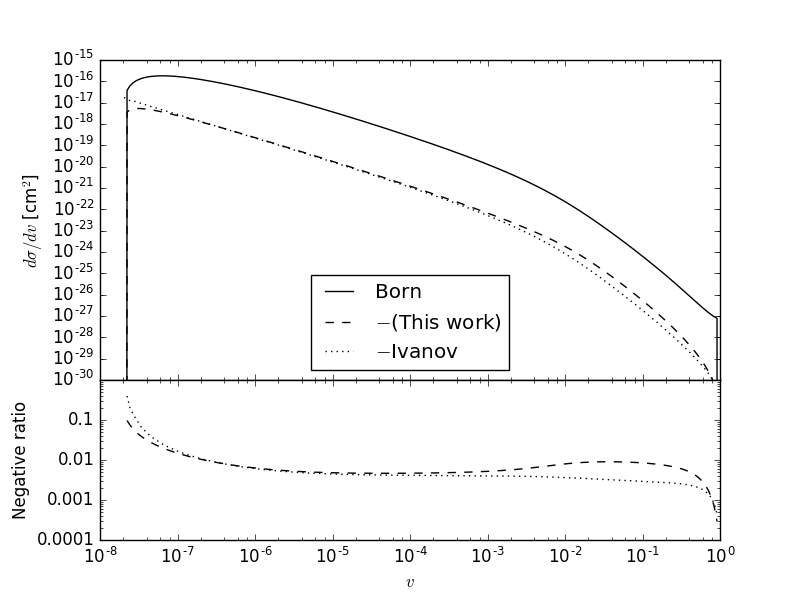}
\caption{Differential cross section $d\sigma/dv$ for a muon of \SI{100}{TeV}
primary energy in standard rock. Shown are the cross section in Born
approximation \cite{KokoulinPetrukhin2} (solid line), our Coulomb corrections
(dashed line), and the corrections of \cite{Melnikov} (dotted line).}
\label{diff-standard-rock}
\end{figure}

\begin{figure}
\includegraphics[width=\textwidth]{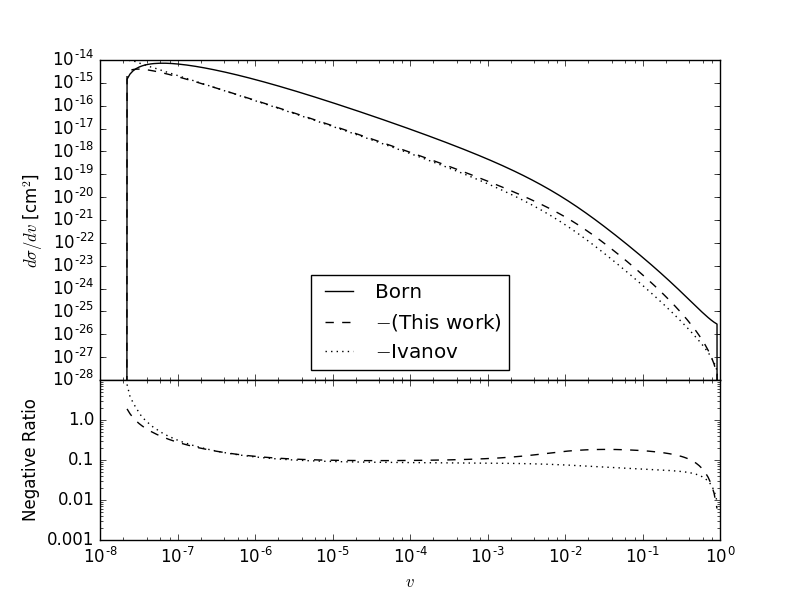}
\caption{Differential cross section $d\sigma/dv$ for a muon of \SI{100}{TeV}
primary energy in lead. Shown are the cross section in Born approximation
\cite{KokoulinPetrukhin2} (solid line), our Coulomb corrections (dashed line),
and the corrections of \cite{Melnikov} (dotted line).}
\label{diff-lead}
\end{figure}

Using the result for the process of pair production by a virtual photon, one can
calculate the Coulomb corrections to the cross section of pair production by a
charged particle, thus generalizing the corrections calculated by
\cite{Melnikov} for pair production in a Coulomb field. The correction to the
cross section for pair production by a muon is given by
\begin{equation}
\diff{\sigma_2} = \diff{n_T}(\omega, Q^2) \sigma_2^T(\omega, Q^2)
  + \diff{n_S} (\omega, Q^2) \sigma_2^S(\omega, Q^2),
\end{equation}
where the virtual photon fluxes are given by \cite{Budnev}
\begin{align}
\diff{n_T}(\omega, Q^2) &= \frac{\alpha}{\pi} (1 - v) \left(1 -
  \frac{Q^2_\text{min}}{Q^2} + \frac{v^2}{2 (1 - v)}\right) \frac{\diff\omega}
  {\omega} \frac{\diff{Q^2}}{Q^2},\\
\diff{n_S}(\omega, Q^2) &= \frac{\alpha}{\pi} (1 - v) \frac{\diff\omega}
  {\omega} \frac{\diff{Q^2}}{Q^2},\\
Q^2_\text{min} = \frac{m_\mu^2 v^2}{1 - v} &\leq Q^2 < \infty.
\end{align}
where $v = \omega/E_\mu$ is the fractional energy loss of the muon. Since
$\sigma_2^T, \sigma_2^S$ are independent of $\omega$ and $\diff{n_T,}
\diff{n_S,} Q^2_\text{min}$ only depend on the fractional energy loss, the
correction itself is independent of the incident muon energy, because the
singularity for $x \rightarrow 0, x \rightarrow 1$ is only logarithmic and
therefore integrable. Since the contribution in Born approximation is dependent
on energy, however, the relative importance of the correction is a function of
the energy. Also, the integration over $x$ should only be carried out in the
range where the Born contribution is non-negative. The influence of
Coulomb corrections on the differential cross section $d\sigma/dv$ for a muon
of \SI{100}{TeV} primary energy in standard rock and lead is shown in
\figref{diff-standard-rock}, \ref{diff-lead} in comparison to the Born
contribution of \cite{KokoulinPetrukhin2}.

\begin{figure}
\includegraphics[width=\textwidth]{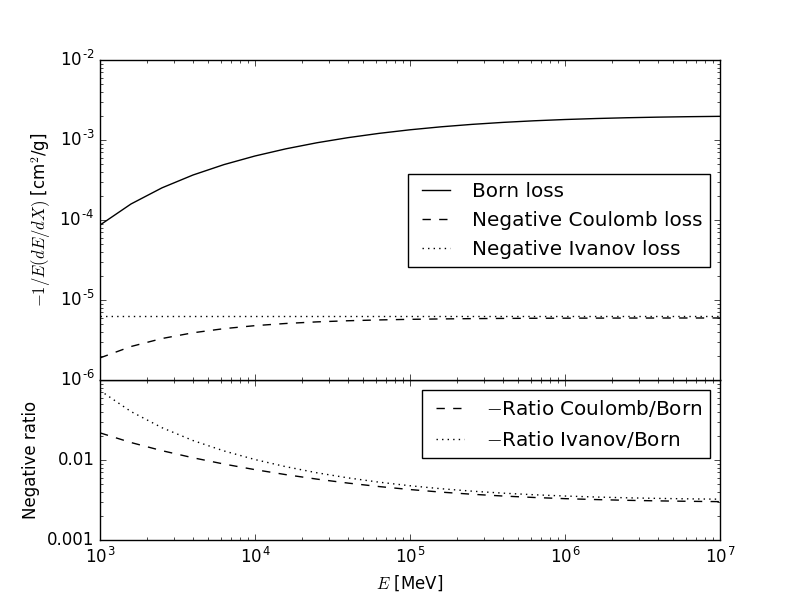}
\caption{Average energy loss through pair production in standard rock,
calculated using the Born cross section of \cite{KokoulinPetrukhin2} (solid
line) and the negative Coulomb corrections calculated in this work (dashed line)
and in \cite{Melnikov} (dotted line).}
\label{coulomb-energy-loss}
\end{figure}

\begin{figure}
\includegraphics[width=\textwidth]{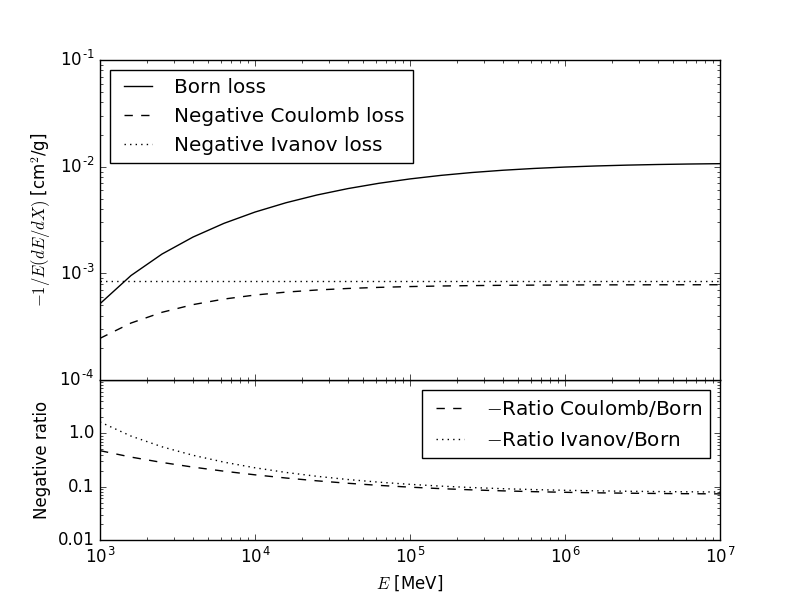}
\caption{Average energy loss through pair production in lead, calculated
using the Born cross section of \cite{KokoulinPetrukhin2} (solid line) and the
negative Coulomb corrections calculated in this work (dashed line) and in
\cite{Melnikov} (dotted line).}
\label{lead-loss}
\end{figure}

The influence of Coulomb corrections on the average energy loss
\begin{equation}
-\frac{\diff E}{\diff X} = \frac{N_A}{A}
  \Int \omega \frac{\diff\sigma}{\diff\omega} \diff\omega,
\end{equation}
where $N_A$ is Avogadro's constant, $A$ is the mass number of the material, and
$X = x/\rho$ is the depth, is shown in \figref{coulomb-energy-loss} for standard
rock\footnote{Standard
rock is assumed as a mixture of MgCO$_3$ and CaCO$_3$ consisting of 52\% oxygen,
27\% calcium and 9\% magnesium.} and in \figref{lead-loss} for lead, integrated
in the appropriate energy-dependent limits of the Born approximation cross
section of \cite{KokoulinPetrukhin2}.

\section{Discussion}
We have calculated Coulomb corrections to the cross sections of pair production
and bremsstrahlung on extended screened nuclei. These calculations generalize
the work of \cite{Ivanov,Melnikov,Bakmaev} with regard to pair production and
the work of \cite{ABB2} with regard to muon bremsstrahlung.

Our results confirm that the Coulomb corrections to the muon bremsstrahlung
cross section are negligible with very high accuracy. This coincides
qualitatively with the results of \cite{ABB2}, who applied a very simple model
for the charge distribution of the nucleus and used a different method based on
wave functions. However, here a more realistic charge distribution was used;
therefore a direct comparison of the numerical results is difficult.
In contrast to the results of \cite{KrachkovBrems}, the corrections do not
vanish identically in our calculation.

Our results on electron pair production by high-energy muons confirm the
importance of Coulomb corrections established by \cite{Melnikov} for this
process in precise calculations of muon transport. Our calculations differ in
two aspects from \cite{Melnikov}:
\begin{itemize}
\item the correction in the cross section is integrated only over values of 
$x, v$, for which the Born cross section is positive;
\item the atomic and nuclear form factor is taken into account.
\end{itemize}
The effect of the first aspect decreases with energy; however, as shown in
\figref{diff-standard-rock}, the effect of correct limits is still noticeable
at a muon energy of \SI{100}{TeV} in standard rock. The second point leads to
an additional decrease of the Coulomb correction which does not decrease with
energy. As shown in \figref{lead-loss}, for lead the correction to the energy
loss is smaller by more than 10\%, amounting to about a percent of the Born
loss. For the differential cross section, the effect is even greater, as shown
in \figref{diff-lead}.

\section*{Acknowledgments}
A.~S. and W.~R. acknowledge funding by the Deutsche Forschungsgemeinschaft
 under the grant number RH~35/9-1. The authors thank
Dmitry Ivanov for explanations of intermediate steps of the original
calculation and Anatoly Petrukhin, Rostislav Kokoulin, Stanislav Kelner
and Jan Soedingrekso for helpful discussions, and Anthony Flores for diligent
proofreading.

\bibliography{references.bib}

\begin{thebibliography}{10}
\expandafter\ifx\csname url\endcsname\relax
  \def\url#1{\texttt{#1}}\fi
\expandafter\ifx\csname urlprefix\endcsname\relax\def\urlprefix{URL }\fi
\expandafter\ifx\csname href\endcsname\relax
  \def\href#1#2{#2} \def\path#1{#1}\fi

\bibitem{KokoulinPetrukhin1}
R.~P. Kokoulin, A.~A. Petrukhin, Analysis of the cross section of direct pair
  production by fast muons, in: Proc. 11th Int. Conf. on Cosmic Rays, Budapest
  1969, Vol. 29, Suppl. 4, Acta Phys. Acad. Sci. Hung., 1970, pp. 277--284.

\bibitem{KokoulinPetrukhin2}
R.~P. Kokoulin, A.~A. Petrukhin, Influence of the nuclear formfactor on the
  cross section of electron pair production by high-energy muons, in: Proc.
  12th Int. Conf. on Cosmic Rays, Hobart 1971, Vol.~6, 1971, pp. 2436--2444.

\bibitem{Kelner-atomic}
S.~R. Kelner, Pair production in collisions between muons and atomic electrons,
  Phys. At. Nucl. 61 (1998) 448--456.

\bibitem{KKP}
S.~R. Kelner, R.~P. Kokoulin, A.~A. Petrukhin, About cross section for
  high-energy muon bremsstrahlung, Preprint MEPhI 024-95, Moscow (1995).

\bibitem{KKP-atomic}
S.~R. Kelner, R.~P. Kokoulin, A.~A. Petrukhin, Bremsstrahlung from muons
  scattered by atomic electrons, Phys. At. Nucl. 60 (1997) 576--583.

\bibitem{ALLM91}
H.~Abramowicz, E.~M. Levin, A.~Levy, U.~Maor, A parametrization of
  $\sigma_t(\gamma^* p)$ above the resonance region for {$Q^2 \geq 0$}, Phys.
  Lett. B 269 (1991) 465--476.
\newblock \href {http://dx.doi.org/10.1016/0370-2693(91)90202-2}
  {\path{doi:10.1016/0370-2693(91)90202-2}}.

\bibitem{ALLM97}
H.~Abramowicz, A.~Levy, The {ALLM} parametrization of
  $\sigma_\text{tot}(\gamma^* p)$: an update, arXiv:hep-ph/9712415 (1997).

\bibitem{BezrukovBugaev}
L.~B. Bezrukov, E.~V. Bugaev, Nucleon shadowing effects in photonuclear
  interactions, Sov. J. Nucl. Phys. 33 (1981) 635.

\bibitem{KokoulinUncertainties}
R.~P. Kokoulin, Uncertainties in underground muon flux calculations, Nucl.
  Phys. B Proc. Suppl. 70 (1999) 475.

\bibitem{PetrukhinShestakov}
A.~A. Petrukhin, V.~V. Shestakov, The influence of nuclear and atomic form
  factors on the muon bremsstrahlung cross section, Canad. J. Phys. 46 (1968)
  S377.

\bibitem{Ivanov}
D.~Ivanov, K.~Melnikov, Lepton pair production by a high energy photon in a
  strong electromagnetic field, Phys. Rev. D 57 (1998) 4025.

\bibitem{Melnikov}
D.~Ivanov, E.~A. Kuraev, A.~Schiller, V.~G. Serbo, Production of $e^+e^-$ pairs
  to all orders in {$Z\alpha$} for collisions of high-energy muons with heavy
  nuclei, Phys. Lett. B 442 (1998) 453--458.

\bibitem{KrachkovPair}
P.~A. Krachkov, A.~I. Milstein, Coulomb effects in high-energy $e^+e^-$
  electroproduction by a heavy charged particles in an atomic field, Phys.
  Lett. B 771 (2017) 5--8.

\bibitem{DaviesBetheMaximon}
H.~A. Bethe, L.~C. Maximon, Theory of bremsstrahlung and pair production. i.
  differential cross section, Phys. Rev. 93 (1954) 768.

\bibitem{ABB2}
Y.~M. Andreev, E.~V. Bugaev, Muon bremsstrahlung on heavy atoms, Phys. Rev. D
  55 (1997) 1233--1243.

\bibitem{KrachkovBrems}
P.~A. Krachkov, A.~I. Milstein, Charge asymmetry in the differential cross
  section of high-energy bremsstrahlung in the field of a heavy atom, Phys.
  Rev. A 91 (2015) 032106.

\bibitem{Bakmaev}
S.~Bakmaev, E.~A. Kuraev, I.~Shapoval, Y.~P. Peresun'ko, Electron-positron pair
  production by linearly polarized photon in the nuclear field, Phys. Lett. B
  660 (2008) 494--500.

\bibitem{Sudakov}
V.~V. Sudakov, Vertex parts at very high energies in quantum electrodynamics,
  Sov. Phys. JETP 3 (1956) 65--71.

\bibitem{Lipatov}
L.~N. Lipatov, G.~V. Frolov, Some processes in quantum electrodynamics at high
  energies, Sov. J. Nucl. Phys. 13 (1971) 333--339.

\bibitem{Ivanov-mail}
D.~Y. Ivanov, Private communications.

\bibitem{Suura}
R.~Yennie, S.~C. Frautschi, H.~Suura, Ann. Phys. (N. Y.) 13 (1961) 379.

\bibitem{Tsai}
Y.~S. Tsai, Pair production and bremsstrahlung of charged leptons, Rev. Mod.
  Phys. 46 (1974) 815--851.

\bibitem{Olsen55}
H.~Olsen, Outgoing and ingoing waves in final states and bremsstrahlung, Phys.
  Rev. 99 (1955) 1335.

\bibitem{Budnev}
V.~M. Budnev, I.~F. Ginzburg, G.~V. Meledin, V.~G. Serbo, The two-photon
  particle production mechanism. {P}hysical problems. {A}pplications.
  {E}quivalent photon approximation, Phys. Rep. 3 (1975) 181--282.

\end{thebibliography}
\end{document}